\documentclass[twocolumn,aps,pra,10pt,superscriptaddress]{revtex4-2}

\usepackage{amsmath, amssymb, amsthm}
\usepackage{physics}
\usepackage{graphicx}
\usepackage{bm}
\usepackage[colorlinks=true, allcolors=blue]{hyperref}
\usepackage{algpseudocode}
\usepackage{mdframed}
\usepackage{caption}
\usepackage{algorithm}          % 提供 algorithm 浮动环境
\floatstyle{ruled}              % 设置浮动体样式：标题上下各一条横线
\restylefloat{algorithm}        % 将样式应用给 algorithm 环境

\begin{document}

\title{Geometric Construction of Optimal Teleportation Witnesses}

\author{Yanning Jia}
\author{Fenzhuo Guo}\email{gfenzhuo@bupt.edu.cn}
\author{Mengxuan Bai}
\author{Mengyan Li}
\affiliation{School of Mathematical Sciences, Beijing University of Posts and Telecommunications, Beijing 100876, China}
\affiliation{Key Laboratory of Mathematics and Information Networks, Beijing University of Posts and Telecommunications, Ministry of Education, China}
\affiliation{State Key Laboratory of Networking and Switching Technology, Beijing University of Posts and Telecommunications, Beijing 100876, China}

\author{Haifeng Dong}
\affiliation{School of Instrumentation Science and Opto-Electronics Engineering, Beihang University, Beijing 100191, China}

\author{Fei Gao}
\affiliation{State Key Laboratory of Networking and Switching Technology, Beijing University of Posts and Telecommunications, Beijing 100876, China}

\begin{abstract}
Not all entangled states are useful for quantum teleportation. We present a geometric method to construct optimal teleportation witnesses, which provide operational necessary and sufficient criteria for identifying the teleportation usefulness of arbitrary two-qudit entangled states. Specifically, we formulate the shortest-distance problem from the target state $\rho$ to the convex set $S$ of useless states and develop a two-layer iterative cutting-plane algorithm to numerically approximate the projection point $\sigma^*\in S$. The corresponding optimal teleportation witness is then constructed from the projection geometry. Moreover, the exact shortest distance $D(\rho)$ provides a necessary and sufficient criterion for usefulness. We apply our method to identify the teleportation usefulness of four classes of entangled states.
\end{abstract}

\maketitle
% =========================
\section{Introduction}
%量子隐形传态的定义。引出判定问题
Quantum teleportation is a fundamental task that allows an unknown quantum state $|\psi\rangle$ to be transmitted from one party to another using a shared entangled state $\rho$ and classical communication \cite{PhysRevLett.70.1895, Bouwmeester1997,PhysRevLett.80.1121}. The performance of teleportation is typically quantified by the average fidelity
\begin{equation}
F(\rho) = \int \langle \psi | \Lambda_\rho(|\psi\rangle\langle\psi|) | \psi \rangle \, d\psi,
\end{equation}
where $\Lambda_\rho$ denotes the optimized teleportation channel implemented using the entangled state $\rho$, and the integral is over the Haar measure of all unknown quantum states \cite{PhysRevA.60.1888, Hu2023}. For a $d$-dimensional system, when the two parties share no entanglement, the average fidelity of teleportation via classical channel is bounded by $F_{\mathrm{cl}} = \frac{2}{d+1}$ \cite{PhysRevA.60.1888}. Accordingly, a quantum advantage is realized only when the teleportation fidelity $F(\rho)$ mediated by the state $\rho$ surpasses $F_{\mathrm{cl}}$. However, not all entangled states can provide a quantum advantage in teleportation \cite{PhysRevLett.72.797,Chakrabarty2010,PhysRevLett.84.4236, Adhikari2010}, making it crucial to efficiently identify whether a given entangled state is useful for teleportation.
%已有的方法和缺陷。

Existing criteria for identifying teleportation usefulness of given states can be roughly divided into two categories. The first one assesses teleportation capability via Bell nonlocality \cite{RevModPhys.86.419} and quantum steering \cite{RevModPhys.92.015001}. For instance, any two‑qubit state violating the CHSH inequality or certain linear steering inequalities is guaranteed to be useful for teleportation \cite{HORODECKI1995340, HORODECKI199621, PhysRevA.106.012433}. In 2025, Guo \textit{et al.} further proposed the average-correlation inequality, the maximal violation of which is closely related to those of CHSH and linear steering inequalities, as a sufficient criterion for identifying useful two-qubit states \cite{lmdx-dk95}. However, all these inequality-based criteria are restricted to two-qubit states and provide only sufficient conditions. The second approach constructs teleportation witnesses based on the maximal entangled fraction (MEF) to assess whether the states are useful for teleportation \cite{PhysRevLett.107.270501, PhysRevA.86.032315}. Existing results show that a two-qudit state is useful for teleportation if and only if $\mathrm{MEF}>1/d$ \cite{PhysRevA.60.1888}. However, since the MEF is generally difficult to evaluate except for a few special cases \cite{Zhao_2010,GuRui-Juan_2010}, these witnesses are usually constructed from computable lower bounds of the MEF and therefore also provide only sufficient conditions for identifying useful two-qudit states. In addition, there exist entangled states that cannot be identified by either of the above two approaches, despite exhibiting quantum advantage in teleportation \cite{Chakrabarty2010, Adhikari2010, PhysRevLett.107.270501}.  Although Ref.~\cite{PhysRevA.85.054301} established necessary and sufficient teleportation-witness criteria for two-qudit states, it did not provide a target-state-specific closest-point construction of the projection-defined optimal witness considered here. Consequently, establishing a systematic and constructive framework for identifying the teleportation usefulness of general two-qudit states remains a motivated direction.

%本文的工作（用到的方法）
In this work, we propose a geometric approach to construct optimal teleportation witnesses for arbitrary two-qudit states. At the level of the exact geometric formulation, this approach provides a necessary and sufficient criterion to determine whether a given state $\rho$ is useful for teleportation. The set of states useless for teleportation forms a convex set $S$ characterized by $\mathrm{MEF}\le 1/d$ \cite{PhysRevA.60.1888, PhysRevLett.107.270501}. Based on the projection theorem \cite{rockafellar1970}, we construct the optimal teleportation witness from the projection geometry at the unique closest state $\sigma^* \in S$ to a given state $\rho$. The main challenge in the construction is finding this closest state $\sigma^*$, which we recast as the solution of a shortest distance optimization problem between $\rho$ and the convex set $S$. This problem is nontrivial because $S$ is defined by uncountably many local-unitary constraints, leading to a semi-infinite programming formulation. We develop a two-layer iterative cutting-plane algorithm to solve it. The outer layer computes a candidate state by solving a relaxed problem with finitely many constraints, while the inner layer numerically searches for a violated unitary constraint to refine the feasible region. This systematic alternation provides a numerical approximation to
$\sigma^*$, from which the corresponding teleportation witness can be
explicitly constructed.  Moreover, the geometric construction process also provides an independent criterion, confirming that $\rho$ is useful for teleportation if and only if the shortest distance $D(\rho)>0$. 

%应用到什么态上的例子
To demonstrate the practical applicability of our method, we test it on
four classes of states. For two-qubit systems, we first consider the
Werner states, for which our method reproduces the known
teleportation-usefulness threshold $p=1/3$ and shows that the resulting
geometrically optimal witness coincides with the teleportation witness
proposed in Ref.~\cite{PhysRevLett.107.270501}. We then examine the
maximally entangled mixed states proposed in
Ref.~\cite{PhysRevA.62.022310}, where our method yields the exact
usefulness threshold, which is not captured by the existing
inequality-based criteria in Ref.~\cite{lmdx-dk95}. Moreover, we provide
the first explicit construction of the optimal teleportation witness for
this family and show that it gives a necessary and sufficient criterion
for teleportation usefulness. For two-qutrit systems, we explore a class
of asymmetric nonmaximally entangled states mixed with a separable state.
The numerically obtained threshold coincides with the analytical upper
bound $p_c\leq0.375$ derived from the MEF, and we further analyze how the
optimal teleportation witness varies with the state parameter $p$. For higher-dimensional systems, we consider special generalized Werner states with local dimensions $d=4,\ldots,8$, introduced in Ref.~\cite{Deng2009SGWS}. Our method obtains the exact analytical usefulness thresholds and identifies a common form for the corresponding optimal teleportation witnesses near their respective thresholds.

\section{Distance-Based Geometric Construction of Optimal Teleportation Witnesses}

A bipartite state is useful for teleportation if it achieves an average
fidelity beyond that attainable by the optimal classical protocol
\cite{PhysRevA.60.1888}. For a bipartite state
$\rho\in\mathcal{D}(\mathbb{C}^d\otimes\mathbb{C}^d)$, where
$\mathcal{D}(\mathcal{H})$ denotes the set of density operators on
$\mathcal{H}$, the optimal average teleportation fidelity is determined
by the maximal entangled fraction $F_{\max}(\rho)$ through
\begin{equation}
F(\rho)
=
\frac{dF_{\max}(\rho)+1}{d+1}.
\label{eq:fidelity}
\end{equation}

Here,
\begin{equation}
\begin{aligned}
F_{\max}(\rho)
&=
\max_{U\in\mathrm{U}(d)}
\operatorname{Tr}(\rho P_U),\\
P_U
&=
(I_d\otimes U)
\ket{\Phi_d^+}\bra{\Phi_d^+}
(I_d\otimes U^\dagger),
\end{aligned}
\label{eq:max_fraction}
\end{equation}
where $\mathrm{U}(d)$ denotes the group of unitary operators on
$\mathbb{C}^d$, $I_d$ is the identity operator on $\mathbb{C}^d$, and
$\ket{\Phi_d^+}=\frac{1}{\sqrt d}\sum_{i=0}^{d-1}\ket{ii}$ is the
$d$-dimensional maximally entangled state
\cite{PhysRevA.60.1888,PhysRevA.62.012304,Zhao_2010}.
The optimal classical teleportation fidelity for a $d$-dimensional
system is $2/(d+1)$. Therefore, using Eq.~\eqref{eq:fidelity}, $\rho$
provides a quantum advantage in teleportation if and only if
\begin{equation}
F_{\max}(\rho)>\frac{1}{d}.
\label{eq:teleportation_condition}
\end{equation}

Accordingly, the set of states that are not useful for teleportation is
defined as \cite{PhysRevLett.107.270501}
\begin{equation}
S
=
\left\{
\sigma\in
\mathcal{D}(\mathbb{C}^d\otimes\mathbb{C}^d)
\,\middle|\,
F_{\max}(\sigma)\leq\frac{1}{d}
\right\}.
\label{eq:set_S}
\end{equation}

Using Eq.~\eqref{eq:max_fraction}, this set can equivalently be written
as
\begin{equation}
S
=
\left\{
\sigma\in
\mathcal{D}(\mathbb{C}^d\otimes\mathbb{C}^d)
\,\middle|\,
\operatorname{Tr}(\sigma P_U)\leq\frac{1}{d},
\quad
\forall U\in\mathrm{U}(d)
\right\}.
\label{eq:set_S_equiv}
\end{equation}

For each fixed $U$, the inequality
$\operatorname{Tr}(\sigma P_U)\leq1/d$ defines a closed convex
half-space in the real vector space of Hermitian operators. Hence, $S$
is the intersection of these half-spaces with the set of density
operators and is therefore closed and convex. It is also nonempty,
since the maximally mixed state $I/d^2$ belongs to $S$, where $I$
denotes the identity operator on
$\mathbb{C}^d\otimes\mathbb{C}^d$. Moreover, as a closed subset of the
compact set of density operators in finite dimensions, $S$ is compact.

For an arbitrary target state $\rho$, we define its Frobenius distance
from $S$ as
\begin{equation}
D(\rho)
=
\min_{\sigma\in S}
\|\rho-\sigma\|_F,
\label{eq:distance}
\end{equation}
where $\|A\|_F=\sqrt{\operatorname{Tr}(A^\dagger A)}$ is the
Frobenius, or Hilbert--Schmidt, norm. For $\rho\notin S$, the
projection theorem for nonempty closed convex sets guarantees the
existence of a unique closest point $\sigma^*\in S$
\cite{rockafellar1970}, satisfying
\begin{equation}
\sigma^*
=
\arg\min_{\sigma\in S}
\|\rho-\sigma\|_F,
\qquad
D(\rho)
=
\|\rho-\sigma^*\|_F.
\label{eq:projection}
\end{equation}

For an arbitrary $\sigma\in S$, convexity implies that
$\sigma_t=\sigma^*+t(\sigma-\sigma^*)$, with $t\in[0,1]$, also belongs
to $S$. Define $g(t)=\|\rho-\sigma_t\|_F^2$. The minimality of
$\sigma^*$ implies $g'_+(0)\geq0$. Direct differentiation therefore
gives
\begin{equation}
\operatorname{Tr}
[(\sigma^*-\rho)(\sigma-\sigma^*)]
\geq0,
\qquad
\forall\sigma\in S.
\label{eq:projection_property}
\end{equation}

\begin{figure}[htbp]
    \centering
    \includegraphics[width=0.45\textwidth]{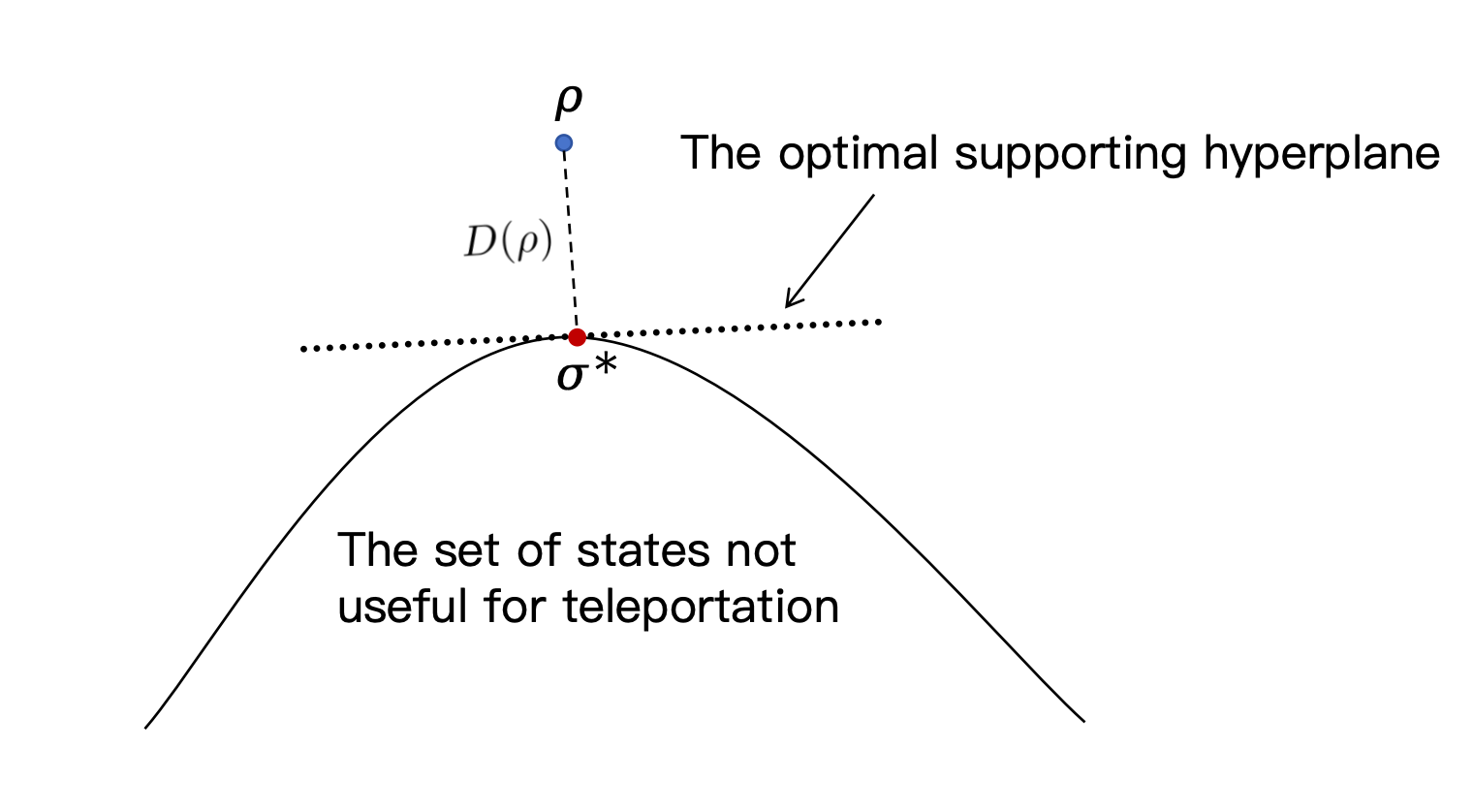}
    \caption{Geometric characterization. The set $S$ of states not useful
    for teleportation is convex. For $\rho\notin S$, the closest point
    $\sigma^*$ defines the distance $D(\rho)$. The supporting hyperplane
    at $\sigma^*$ with normal $n=\sigma^*-\rho$ separates $\rho$ from
    $S$, and the corresponding teleportation witness is obtained from
    this supporting geometry.}
    \label{fig1}
\end{figure}

Let $n=\sigma^*-\rho$. Since $\rho\notin S$, we have $n\neq0$.
Equation~\eqref{eq:projection_property} shows that $S$ lies entirely in
the half-space $\operatorname{Tr}[n(X-\sigma^*)]\geq0$. Hence,
\begin{equation}
\operatorname{Tr}[n(X-\sigma^*)]=0
\label{eq:supporting_hyperplane}
\end{equation}
defines a supporting hyperplane of $S$ at $\sigma^*$, where $X$ is an
arbitrary Hermitian operator on
$\mathbb{C}^d\otimes\mathbb{C}^d$
\cite{rockafellar1970,rudin1991}. For the target state,
$\operatorname{Tr}[n(\rho-\sigma^*)]
=-\|\rho-\sigma^*\|_F^2<0$, so $\rho$ lies strictly on the opposite
side of the supporting hyperplane from $S$. This closest-point geometry
is illustrated in Fig.~\ref{fig1}.

The supporting hyperplane in Eq.~\eqref{eq:supporting_hyperplane}
naturally determines a teleportation witness. Since every density
operator $X$ satisfies $\operatorname{Tr}(X)=1$, define $W=n-cI$, where
$c=\operatorname{Tr}(n\sigma^*)$ and $I$ denotes the identity operator
on $\mathbb{C}^d\otimes\mathbb{C}^d$. Then, for every density operator
$X$,
\begin{equation}
\begin{aligned}
\operatorname{Tr}(WX)
&=
\operatorname{Tr}(nX)
-
c\operatorname{Tr}(X)\\
&=
\operatorname{Tr}(nX)
-
\operatorname{Tr}(n\sigma^*)\\
&=
\operatorname{Tr}[n(X-\sigma^*)].
\end{aligned}
\label{eq:witness_hyperplane}
\end{equation}
Therefore, when restricted to the state space, the condition
$\operatorname{Tr}(WX)=0$ represents exactly the supporting hyperplane
in Eq.~\eqref{eq:supporting_hyperplane}. Using $n=\sigma^*-\rho$, the
witness takes the explicit form
\begin{equation}
W
=
\sigma^*-\rho
-
\operatorname{Tr}
[(\sigma^*-\rho)\sigma^*]I.
\label{eq:witness}
\end{equation}

For any $\sigma\in S$,
\begin{equation}
\operatorname{Tr}(W\sigma)
=
\operatorname{Tr}
[(\sigma^*-\rho)(\sigma-\sigma^*)]
\geq0,
\label{eq:witness_positive}
\end{equation}
where the inequality follows from
Eq.~\eqref{eq:projection_property}. In particular,
$\operatorname{Tr}(W\sigma^*)=0$. For the target state,
\begin{equation}
\operatorname{Tr}(W\rho)
=
-\|\rho-\sigma^*\|_F^2
=
-D(\rho)^2
<0.
\label{eq:witness_violation}
\end{equation}
Therefore, $W$ is a valid teleportation witness: it is nonnegative on
all states in $S$ and negative on $\rho$.

Throughout this work, ``optimal'' refers specifically to the
Hilbert--Schmidt nearest-point construction: the witness in
Eq.~\eqref{eq:witness} is selected by the unique closest point
$\sigma^*$ of $\rho$ in $S$, with its supporting direction fixed by
$\sigma^*-\rho$. This notion is analogous to the nearest-point
construction used for entanglement witnesses
\cite{PhysRevA.66.032319,PhysRevA.72.052331}.

Since $S$ contains exactly the states that are not useful for
teleportation,
\begin{equation}
D(\rho)>0
\quad\Longleftrightarrow\quad
\rho\notin S
\quad\Longleftrightarrow\quad
F_{\max}(\rho)>\frac{1}{d}.
\label{eq:distance_criterion}
\end{equation}
Thus, $D(\rho)>0$ is necessary and sufficient for $\rho$ to be useful
for teleportation. Together with Eq.~\eqref{eq:witness_violation}, this
shows that solving the shortest-distance problem simultaneously
determines the distance $D(\rho)$, the closest point $\sigma^*$, the
supporting hyperplane, and the corresponding optimal teleportation
witness.

Using the characterization of $S$ in Eq.~\eqref{eq:set_S_equiv}, the
shortest-distance problem can be written explicitly as
\begin{equation}
\begin{aligned}
\min_{\sigma}\quad
&\|\rho-\sigma\|_F\\
\mathrm{s.t.}\quad
&\sigma\succeq0,\qquad
\operatorname{Tr}(\sigma)=1,\\
&\operatorname{Tr}(\sigma P_U)\leq\frac{1}{d},
\qquad
\forall U\in\mathrm{U}(d).
\end{aligned}
\label{eq:primal}
\end{equation}

The objective and all constraints are convex. Since the last family of
constraints is indexed by the continuous unitary group
$\mathrm{U}(d)$, Eq.~\eqref{eq:primal} is a convex semi infinite
optimization problem. Although this family
contains infinitely many constraints, each constraint is linear in
$\sigma$ and corresponds to a half space defining $S$. This structure
naturally suggests using a cutting plane method to approximate $S$.
A finite subset of these half spaces gives an outer approximation of
$S$, which can then be progressively tightened by adding violated
constraints.

Cutting plane methods have previously been used in entanglement witness
search \cite{Ioannou2004,PhysRevA.73.052314}. In these works, separable
states obtained through global optimization are used to generate cutting
planes, which progressively refine an outer approximation of the feasible
region for the witness coefficients until a feasible entanglement witness is obtained. In our method, the half spaces
parameterized by unitary operators that define $S$ are used as cutting
planes to progressively refine an outer approximation of the
teleportation useless set $S$ in the state space. Solving the shortest
distance problem over these approximations yields the closest point
$\sigma^*$, from whose supporting hyperplane the corresponding optimal
teleportation witness is constructed.

\section{Two-Layer Cutting-Plane Method for Distance Computation}

Following the above construction, we implement the cutting-plane
procedure through two coupled optimization layers. At each outer
iteration, a finite subset of the unitary constraints is retained to
form a convex outer approximation of $S$. The resulting projection
problem produces a candidate state $\sigma^{(k)}$, after which an inner
unitary optimization searches for a violated constraint. If a violation
exceeding the prescribed tolerance is found, the corresponding unitary
constraint is added and the outer problem is solved again.

The two layers are therefore:

\begin{itemize}
    \item \textbf{Outer layer:} solve a finite-constraint relaxation of
    Eq.~\eqref{eq:primal} and obtain the projection $\sigma^{(k)}$ onto
    the current outer approximation $S_k$.

    \item \textbf{Inner layer:} for $\sigma^{(k)}$, numerically maximize
    $\operatorname{Tr}(\sigma^{(k)}P_U)$ over $U\in\mathrm{U}(d)$ and
    search for a violated unitary constraint.
\end{itemize}

The exact-oracle convergence and the finite-accuracy behavior of the
practical implementation are discussed below.

\subsection{Outer layer: finite-constraint projection}

At the $k$th outer iteration, let
$\mathcal{U}_k\subset\mathrm{U}(d)$ denote the current finite active set
of unitary constraints. Since the square function is strictly increasing
on the nonnegative real axis, minimizing the Frobenius distance over a
fixed feasible set is equivalent to minimizing its square. We therefore
solve
\begin{equation}
\begin{aligned}
\sigma^{(k)}
=
\arg\min_{\sigma}\quad
&\frac12\|\rho-\sigma\|_F^2\\
\mathrm{s.t.}\quad
&\sigma\succeq0,\qquad
\operatorname{Tr}(\sigma)=1,\\
&\operatorname{Tr}(\sigma P_U)\leq\frac1d,
\qquad
\forall U\in\mathcal{U}_k.
\end{aligned}
\label{eq:outer}
\end{equation}

This is a finite-dimensional convex quadratic semidefinite program. It
has the same minimizer as
$\min_{\sigma\in S_k}\|\rho-\sigma\|_F$ and constitutes a relaxation of
the original semi-infinite problem in Eq.~\eqref{eq:primal}. We denote
the feasible set of Eq.~\eqref{eq:outer} by $S_k$. Since
$\mathcal{U}_k\subset\mathrm{U}(d)$, every state in the exact
teleportation-useless set $S$ satisfies all constraints defining $S_k$;
hence $S\subseteq S_k$, and $\sigma^{(k)}$ is the closest point to
$\rho$ in this current outer approximation.

Whenever a new valid unitary constraint is added,
$\mathcal{U}_{k+1}\supseteq\mathcal{U}_k$, so the feasible sets satisfy
\begin{equation}
S
\subseteq
S_{k+1}
\subseteq
S_k.
\label{eq:nestedsets}
\end{equation}

The outer approximations therefore become progressively tighter while
always containing the true set $S$.

\subsection{Inner layer: unitary optimization}

For the candidate state $\sigma^{(k)}$, the inner layer considers
\begin{equation}
\begin{aligned}
F_{\max}(\sigma^{(k)})
&=
\max_{U\in\mathrm{U}(d)}F_{\sigma^{(k)}}(U),\\
F_{\sigma^{(k)}}(U)
&=
\operatorname{Tr}(\sigma^{(k)}P_U),
\end{aligned}
\label{eq:innermax}
\end{equation}
where $P_U=(I_d\otimes U)P_0(I_d\otimes U^\dagger)$ and
$P_0=|\Phi_d^+\rangle\langle\Phi_d^+|$. If the global maximum exceeds
$1/d$, the corresponding maximizing unitary specifies a violated
defining constraint of $S$.

To evaluate the objective, write
$\sigma^{(k)}_{ab,ce}=\langle a,b|\sigma^{(k)}|c,e\rangle$. Using
$(I_d\otimes U)|\Phi_d^+\rangle
=\frac{1}{\sqrt d}\sum_{a,b=0}^{d-1}U_{ba}|a,b\rangle$, define the
$d\times d$ matrix $M(U)$ and evaluate the objective as
\begin{equation}
\begin{aligned}
[M(U)]_{ba}
&=
\sum_{c,e=0}^{d-1}
\sigma^{(k)}_{ab,ce}U_{ec},\\
F_{\sigma^{(k)}}(U)
&=
\frac1d
\operatorname{Tr}\!\left[U^\dagger M(U)\right].
\end{aligned}
\label{eq:Mdef}
\end{equation}

Because the optimization in Eq.~\eqref{eq:innermax} is nonconvex over
$\mathrm{U}(d)$, we employ a multi-start polar fixed-point iteration,
following the standard multi-start strategy commonly used in nonconvex
global optimization \cite{RinnooyKan1987}. Specifically, we perform
$R$ independent runs from different initial unitaries
$\{U_0^{(r)}\}_{r=1}^R$. These initial unitaries are chosen
Haar-randomly, with the identity optionally included as one
deterministic initialization.

For the $r$th run, with $r=1,\ldots,R$, let $U_t^{(r)}$ denote the
unitary at the $t$th fixed-point iteration. At iteration $t$, construct
$M_t=M(U_t^{(r)})$ and compute its singular-value decomposition
$M_t=L_t\Sigma_tV_t^\dagger$. The next iterate is chosen as the unitary
polar factor,
\begin{equation}
U_{t+1}^{(r)}
=
L_tV_t^\dagger.
\label{eq:polarupdate}
\end{equation}

The $r$th run is terminated when
\begin{equation}
\left|
F_{\sigma^{(k)}}(U_{t+1}^{(r)})
-
F_{\sigma^{(k)}}(U_t^{(r)})
\right|
\leq
\delta,
\label{eq:innerstop}
\end{equation}
or when $T_{\max}$ inner iterations are reached. Let $F_r$ denote the
final objective value obtained from the $r$th run. After all $R$ runs,
we set $F_{\max}^{\mathrm{est}}=\max_{r=1,\ldots,R}F_r$ and denote the
corresponding numerically best unitary by $U_{\mathrm{best}}$. Since a
finite multi-start search is not guaranteed to attain the global maximum
of Eq.~\eqref{eq:innermax}, we use the notation $U_{\mathrm{best}}$
rather than $U_{\mathrm{opt}}$.

\subsection{Constraint update and termination}

Let $\epsilon\geq0$ denote the constraint-violation tolerance. If
$F_{\max}^{\mathrm{est}}>1/d+\epsilon$, then $U_{\mathrm{best}}$
generates a violated constraint whose violation exceeds the prescribed
tolerance. We update
$\mathcal{U}_{k+1}=\mathcal{U}_k\cup\{U_{\mathrm{best}}\}$ and solve the
outer problem again over the tighter set $S_{k+1}$. If instead
$F_{\max}^{\mathrm{est}}\leq1/d+\epsilon$, the practical algorithm
terminates because the inner search has not detected a violation
exceeding the prescribed tolerance. Since the inner optimization is
nonconvex, this stopping condition is a numerical criterion and, for a
finite multi-start search, should not be interpreted as a proof that the
global maximum satisfies the same inequality. The maximum number of
outer iterations $K_{\max}$ is used only as a safety limit; reaching
this limit is recorded separately from tolerance-based termination.

Every unitary $U\in\mathrm{U}(d)$ defines the valid constraint
$\operatorname{Tr}(\sigma P_U)\leq1/d$ of the exact set $S$.
Consequently, every cutting plane generated by the inner procedure is
valid, irrespective of whether $U_{\mathrm{best}}$ is a global
maximizer. Therefore, the inclusion $S\subseteq S_k$ is preserved
throughout the procedure.

Define
\begin{equation}
D_k
=
\|\rho-\sigma^{(k)}\|_F
=
\min_{\sigma\in S_k}
\|\rho-\sigma\|_F.
\label{eq:Dk}
\end{equation}

Since $S\subseteq S_k$, we have $0\leq D_k\leq D(\rho)$. Thus, for the
exact solution of the finite outer problem, $D_k$ is a lower bound on
the true distance $D(\rho)$. In particular, $D_k>0$ implies
$D(\rho)>0$ and therefore certifies teleportation usefulness.

In direct analogy with the exact witness in Eq.~\eqref{eq:witness}, the
projection onto $S_k$ defines
\begin{equation}
W_k
=
\sigma^{(k)}-\rho
-
\operatorname{Tr}
[(\sigma^{(k)}-\rho)\sigma^{(k)}]I.
\label{eq:Wk}
\end{equation}

Since $\sigma^{(k)}$ is the closest point to $\rho$ in the convex set
$S_k$, $\operatorname{Tr}(W_k\sigma)\geq0$ for every $\sigma\in S_k$,
and hence for every $\sigma\in S$ because $S\subseteq S_k$. Moreover,
$\operatorname{Tr}(W_k\rho)=-D_k^2$. Hence, for the exact finite outer
solution, whenever $D_k>0$, $W_k$ is already a valid teleportation
witness for $\rho$.

At termination, we denote the final iterate by
$\sigma_{\mathrm{out}}=\sigma^{(k)}$, $D_{\mathrm{out}}=D_k$, and
$W_{\mathrm{out}}=W_k$. These are the numerical quantities used in the
applications below. When only the witness direction is relevant, we use
the numerical unit-Frobenius representative
$\widehat W_{\mathrm{out}}=W_{\mathrm{out}}/\|W_{\mathrm{out}}\|_F$ for
$D_{\mathrm{out}}>0$. If a different positive rescaling
$\widetilde W_{\mathrm{out}}=\kappa W_{\mathrm{out}}$, with $\kappa>0$,
is introduced for displaying explicit basis coefficients, it does not
change the witness direction or detection condition.

\par\medskip
\noindent\textbf{Algorithm 1. Practical two-layer iterative cutting-plane method}
\par\smallskip
\begin{algorithmic}[1]

\State \textbf{Input:}
target state $\rho$,
constraint-violation tolerance $\epsilon\geq0$,
inner convergence tolerance $\delta>0$,
number of starts $R\geq1$,
maximum inner iterations $T_{\max}\geq1$,
and maximum outer iterations $K_{\max}\geq1$

\State Initialize
$\mathcal{U}_0=\{I_d\}$ and
\texttt{status}$\leftarrow$\texttt{max-iterations}

\For{$k=0$ to $K_{\max}-1$}

    \State Solve Eq.~\eqref{eq:outer} to obtain $\sigma^{(k)}$
    \State $F_{\mathrm{best}}\leftarrow-\infty$

    \For{$r=1$ to $R$}

        \State Choose an initial unitary $U_0^{(r)}$
        \Statex \hspace{\algorithmicindent}
        \textit{(Haar-random; one start may be chosen as $I_d$)}

        \State
        $F_0^{(r)}
        \leftarrow
        F_{\sigma^{(k)}}(U_0^{(r)})$

        \For{$t=0$ to $T_{\max}-1$}

            \State Construct
            $M_t=M(U_t^{(r)})$
            using Eq.~\eqref{eq:Mdef}

            \State Compute
            $M_t=L_t\Sigma_tV_t^\dagger$

            \State
            $U_{t+1}^{(r)}
            \leftarrow
            L_tV_t^\dagger$

            \State
            $F_{t+1}^{(r)}
            \leftarrow
            F_{\sigma^{(k)}}(U_{t+1}^{(r)})$

            \State
            $\Delta_F
            \leftarrow
            |F_{t+1}^{(r)}-F_t^{(r)}|$

            \If{$\Delta_F\leq\delta$}
                \State \textbf{break}
            \EndIf

        \EndFor

        \State Let $U_{\mathrm{fin}}^{(r)}$ and $F_r$
        denote the final unitary and objective value of the $r$th run

        \If{$F_r>F_{\mathrm{best}}$}

            \State
            $F_{\mathrm{best}}
            \leftarrow
            F_r$

            \State
            $U_{\mathrm{best}}
            \leftarrow
            U_{\mathrm{fin}}^{(r)}$

        \EndIf

    \EndFor

    \State
    $F_{\max}^{\mathrm{est}}
    \leftarrow
    F_{\mathrm{best}}$

    \If{$F_{\max}^{\mathrm{est}}
        \leq1/d+\epsilon$}

        \State
        \texttt{status}
        $\leftarrow$
        \texttt{tolerance-reached}

        \State \textbf{break}

    \ElsIf{$k<K_{\max}-1$}

        \State
        $\mathcal{U}_{k+1}
        \leftarrow
        \mathcal{U}_k
        \cup
        \{U_{\mathrm{best}}\}$

    \Else

        \State \textbf{break}

    \EndIf

\EndFor

\State
$\sigma_{\mathrm{out}}
\leftarrow
\sigma^{(k)}$

\State
$D_{\mathrm{out}}
\leftarrow
\|\rho-\sigma_{\mathrm{out}}\|_F$

\State
$W_{\mathrm{out}}
\leftarrow
\sigma_{\mathrm{out}}-\rho
-
\operatorname{Tr}
[(\sigma_{\mathrm{out}}-\rho)\sigma_{\mathrm{out}}]I$

\State \textbf{Output:}
$\sigma_{\mathrm{out}}$,
$D_{\mathrm{out}}$,
$W_{\mathrm{out}}$,
and \texttt{status}

\end{algorithmic}
\par\medskip

\subsection{Computational complexity}

The computational cost is dominated by the convex quadratic
semidefinite program in the outer layer and the unitary search in the
inner layer.

In the outer problem, $\sigma$ is a $d^2\times d^2$ Hermitian
positive-semidefinite matrix. Let $N=d^2$, $m_k=|\mathcal{U}_k|$, and
$\bar m_k=m_k+1$, where the additional affine constraint in $\bar m_k$
accounts for $\operatorname{Tr}(\sigma)=1$.

Under a representative dense primal-dual interior-point arithmetic
model, the cost of one Newton step for a semidefinite block of size $N$
with $\bar m_k$ affine constraints scales as
$O(\bar m_kN^3+\bar m_k^2N^2+\bar m_k^3)$. The $m_k$ scalar inequality
constraints contribute to the barrier parameter, whereas the equality
constraint $\operatorname{Tr}(\sigma)=1$ does not. Including the
corresponding path-following iteration factor
$O^*(\sqrt{N+m_k})$, where $O^*(\cdot)$ suppresses logarithmic factors
associated with solver accuracy, one outer solve has the representative
complexity
\begin{equation}
\begin{aligned}
C_{\mathrm{outer}}(d,m_k)
&=
O^*\!\left[
\sqrt{d^2+m_k}
\left(
\bar m_kd^6
+\bar m_k^2d^4
+\bar m_k^3
\right)
\right]\\
&=
O^*\!\left(
\bar m_kd^7
+\bar m_k^2d^5
+\bar m_k^3d
\right),
\quad \bar m_k\ll d^2.
\end{aligned}
\label{eq:Couter}
\end{equation}

Thus, for a fixed number of active constraints, the arithmetic cost of
one outer solve scales polynomially with the local dimension $d$, with
leading dependence $O^*(d^7)$ under the adopted dense interior-point
model \cite{doi:10.1137/1038003,BoydVandenberghe2004}.

For the inner layer, constructing $M(U)$ directly from
Eq.~\eqref{eq:Mdef} requires $O(d^4)$ arithmetic operations, whereas the
singular-value decomposition of the resulting $d\times d$ matrix costs
$O(d^3)$. Hence, one fixed-point update is dominated by $O(d^4)$, and
the worst-case arithmetic cost of the $R$-start inner search at one
outer iteration is
\begin{equation}
C_{\mathrm{inner}}
=
O(RT_{\max}d^4).
\label{eq:Cinner}
\end{equation}

If the cutting-plane procedure performs $K\leq K_{\max}$ outer
iterations, the total arithmetic cost is
\begin{equation}
C_{\mathrm{total}}
=
\sum_{k=0}^{K-1}
C_{\mathrm{outer}}(d,m_k)
+
O(KRT_{\max}d^4).
\label{eq:Ctotal}
\end{equation}

For fixed $K$, $R$, $T_{\max}$, and active-set sizes $m_k$, the leading
dimension dependence is therefore $O^*(d^7)$ under this dense-solver
model. This estimate characterizes the arithmetic scaling under fixed
algorithmic budgets and is not a worst-case bound on the number of
cutting-plane iterations required to solve the original semi-infinite
problem to arbitrary accuracy. The outer quadratic semidefinite
optimization is therefore expected to become the principal
computational bottleneck as $d$ increases.

\subsection{Convergence and numerical accuracy}

For each run in the inner layer, the polar fixed-point update is
monotonic in the objective value. Indeed, for
$M_t=M(U_t^{(r)})=L_t\Sigma_tV_t^\dagger$, the unitary polar factor
$L_tV_t^\dagger$ maximizes
$\operatorname{Re}\operatorname{Tr}(Q^\dagger M_t)$ over
$Q\in\mathrm{U}(d)$ \cite{Higham1986Polar}. Moreover, because
$\sigma^{(k)}\succeq0$, the quadratic form associated with the linear
map $M(\cdot)$ is positive semidefinite. The variational property of the
polar factor gives
\begin{equation}
\operatorname{Re}\operatorname{Tr}
\!\left[(U_{t+1}^{(r)})^\dagger M(U_t^{(r)})\right]
\geq
\operatorname{Tr}
\!\left[(U_t^{(r)})^\dagger M(U_t^{(r)})\right].
\label{eq:polar_variational}
\end{equation}

The positive-semidefinite quadratic form further implies, for arbitrary
matrices $U$ and $Q$,
\begin{equation}
F_{\sigma^{(k)}}(Q)
\geq
F_{\sigma^{(k)}}(U)
+\frac{2}{d}
\operatorname{Re}\operatorname{Tr}
\!\left[(Q-U)^\dagger M(U)\right].
\label{eq:quadratic_bound}
\end{equation}
Taking $U=U_t^{(r)}$ and $Q=U_{t+1}^{(r)}$ yields
\begin{equation}
F_{\sigma^{(k)}}(U_{t+1}^{(r)})
\geq
F_{\sigma^{(k)}}(U_t^{(r)}).
\label{eq:innermonotone}
\end{equation}
Since $0\leq F_{\sigma^{(k)}}(U)\leq1$, the objective sequence generated
by each run is monotone and bounded, and therefore converges. This
convergence concerns the objective values along a single fixed-point
run and does not imply convergence to the global maximum of the
nonconvex unitary problem. Using multiple independent starts reduces the
dependence on a particular initialization and improves the practical
reliability of the search \cite{RinnooyKan1987}.

For the outer layer, the nesting relation in
Eq.~\eqref{eq:nestedsets} gives
\begin{equation}
D_k
\leq
D_{k+1}
\leq
D(\rho).
\label{eq:Dmonotone}
\end{equation}

Thus, the sequence $\{D_k\}$ generated before termination is monotone
nondecreasing and bounded above by the true distance.

We first consider an idealized exact-oracle setting. Suppose that each
finite outer problem is solved exactly, the inner maximization is solved
globally, the violation tolerance is set to $\epsilon=0$, and the
procedure is not truncated by $K_{\max}$. The state space and
$\mathrm{U}(d)$ are compact, and the constraint function
$(\sigma,U)\mapsto\operatorname{Tr}(\sigma P_U)-1/d$ is continuous.
Under these conditions, standard convergence results for exchange
methods for infinitely constrained optimization imply that the
procedure either terminates at an exact solution after finitely many
iterations or, if it generates an infinite sequence
\cite{Blankenship1976,doi:10.1137/1035089}, then
$D_k\to D(\rho)$.

All $\sigma^{(k)}$ belong to the compact state space and therefore have
accumulation points. Under the preceding exact-oracle assumptions, every
accumulation point is feasible for the original problem and attains the
limiting distance $D(\rho)$. Since the Frobenius projection onto the
closed convex set $S$ is unique, every accumulation point must equal
$\sigma^*$. It follows that $\sigma^{(k)}\to\sigma^*$, and by the
definition of $W_k$ in Eq.~\eqref{eq:Wk} and continuity in
$\sigma^{(k)}$, also $W_k\to W$, where $W$ is the optimal teleportation
witness defined in Eq.~\eqref{eq:witness}.

We next consider an exact global inner oracle with a finite violation
tolerance $\epsilon>0$. Suppose that the terminating candidate satisfies
$F_{\max}(\sigma^{(k)})\leq1/d+\epsilon$. Define
\begin{equation}
\lambda_\epsilon
=
\frac{\epsilon}
{\frac{d-1}{d^2}+\epsilon},
\qquad
\tau_k
=
(1-\lambda_\epsilon)\sigma^{(k)}
+\lambda_\epsilon\frac{I_{d^2}}{d^2}.
\label{eq:epsmix}
\end{equation}

Since $F_{\max}$ is the pointwise maximum of linear functions of the
state, it is convex, and
$F_{\max}(I_{d^2}/d^2)=1/d^2$. Therefore,
\begin{equation}
F_{\max}(\tau_k)
\leq
(1-\lambda_\epsilon)
\left(\frac1d+\epsilon\right)
+\frac{\lambda_\epsilon}{d^2}
=
\frac1d,
\label{eq:epsfeasible}
\end{equation}
so $\tau_k\in S$. Furthermore,
\begin{equation}
\|\tau_k-\sigma^{(k)}\|_F
\leq
\lambda_\epsilon
\sqrt{1-\frac1{d^2}},
\label{eq:epsdistance}
\end{equation}
where the inequality follows from
$\operatorname{Tr}[(\sigma^{(k)})^2]\leq1$. Hence,
\begin{equation}
\operatorname{dist}(\sigma^{(k)},S)
\leq
\Delta_\epsilon,
\qquad
\Delta_\epsilon
=
\frac{\epsilon}
{\frac{d-1}{d^2}+\epsilon}
\sqrt{1-\frac1{d^2}},
\label{eq:epsbound}
\end{equation}
where $\operatorname{dist}(\cdot,S)$ denotes the Frobenius distance to
$S$. Since $\tau_k\in S$, the triangle inequality gives
$D(\rho)\leq\|\rho-\tau_k\|_F\leq D_k+\Delta_\epsilon$. Together with
$D_k\leq D(\rho)$, this yields
\begin{equation}
0
\leq
D(\rho)-D_k
\leq
\Delta_\epsilon.
\label{eq:Derror}
\end{equation}

In the practical implementation, the finite multi-start polar
fixed-point method provides only a numerical oracle for the inner
optimization. Since Eq.~\eqref{eq:innermax} is nonconvex, the global
maximum is not guaranteed to be attained, and
$F_{\max}^{\mathrm{est}}\leq F_{\max}(\sigma^{(k)})$. Accordingly, the
stopping condition $F_{\max}^{\mathrm{est}}\leq1/d+\epsilon$ should be
interpreted as practical numerical termination under the adopted
multi-start search and prescribed tolerances, rather than as a
certificate of global feasibility of $\sigma^{(k)}$ for the original
set $S$.

Importantly, this limitation does not invalidate the cutting planes that
have already been generated. Every added constraint is a defining
constraint of $S$, so the mathematical inclusion $S\subseteq S_k$ is
preserved independently of the quality of the inner search. Hence, the
exact solution of each finite outer problem provides the lower bound
$D_k\leq D(\rho)$ and, whenever $D_k>0$, a valid teleportation witness
$W_k$. In floating-point calculations, these statements hold up to the
numerical tolerance of the outer SDP solver. A positive distance should
therefore be regarded as numerically resolved only when it exceeds the
corresponding solver accuracy.

\section{Numerical Examples}

In this section, we apply the proposed geometric method to four classes
of bipartite states. For two-qubit systems,
we consider Werner states and a class of maximally entangled mixed
states. For two-qutrit systems, we study an asymmetric nonmaximally
entangled family for which the optimal witness direction varies with the
state parameter. Finally, special generalized Werner states with local
dimensions $d=4,\ldots,8$ are used to examine the method in
higher-dimensional systems.

To keep the witness notation consistent, for any teleportation-useful
target state $\rho$, let $\sigma^*(\rho)$ denote its exact closest point
in the teleportation-useless set $S$. In the numerical calculations,
$\sigma_{\mathrm{out}}[\rho]$ denotes the final numerical approximation
to $\sigma^*(\rho)$ and $W_{\mathrm{out}}[\rho]$ the corresponding
numerical witness. We reserve $W^{\mathrm{geo}}[\rho]$ for the witness
obtained directly from the exact closest-point construction,
\begin{equation}
W^{\mathrm{geo}}[\rho]
=
\sigma^*(\rho)-\rho
-
\operatorname{Tr}
\left\{
[\sigma^*(\rho)-\rho]\sigma^*(\rho)
\right\}I,
\label{eq:example_geo_witness}
\end{equation}
which satisfies
$\operatorname{Tr}[W^{\mathrm{geo}}[\rho]\rho]=-D(\rho)^2<0$.
Since multiplication by a positive constant changes neither the
detection condition nor the supporting-hyperplane direction, we use
the unit-Frobenius representatives
\begin{equation}
\widehat W[\rho]
=
\frac{W^{\mathrm{geo}}[\rho]}{\|W^{\mathrm{geo}}[\rho]\|_F}.
\label{eq:example_normalized_witness}
\end{equation}
For the numerical output, we correspondingly use
\begin{equation}
\widehat W_{\mathrm{out}}[\rho]
=
\frac{W_{\mathrm{out}}[\rho]}{\|W_{\mathrm{out}}[\rho]\|_F}.
\label{eq:example_normalized_witness_out}
\end{equation}

Both representatives have unit Frobenius norm. When a different positive
rescaling is used only to display explicit basis coefficients, its
scaling factor is denoted by $\kappa>0$.

\subsection{Two-Qubit Systems}

We first consider two representative families of two-qubit states.
Werner states provide a case with a known teleportation threshold and a
previously known teleportation-witness form, whereas maximally entangled
mixed states allow comparison with existing sufficient criteria.

\subsubsection{Werner States}

Let $|\Phi^+\rangle=(|00\rangle+|11\rangle)/\sqrt2$. The two-qubit
Werner states considered here are
\begin{equation}
\rho_{\rm W}(p)
=
p|\Phi^+\rangle\langle\Phi^+|
+
(1-p)\frac{I_4}{4},
\qquad
p\in[0,1]
\label{eq:werner_state}
\end{equation}
\cite{PhysRevA.40.4277}. Their maximal entangled fraction is
\begin{equation}
F_{\max}[\rho_{\rm W}(p)]
=
\frac{1+3p}{4}.
\label{eq:werner_mef}
\end{equation}

\begin{figure}[htbp]
    \centering
    \includegraphics[width=\columnwidth]{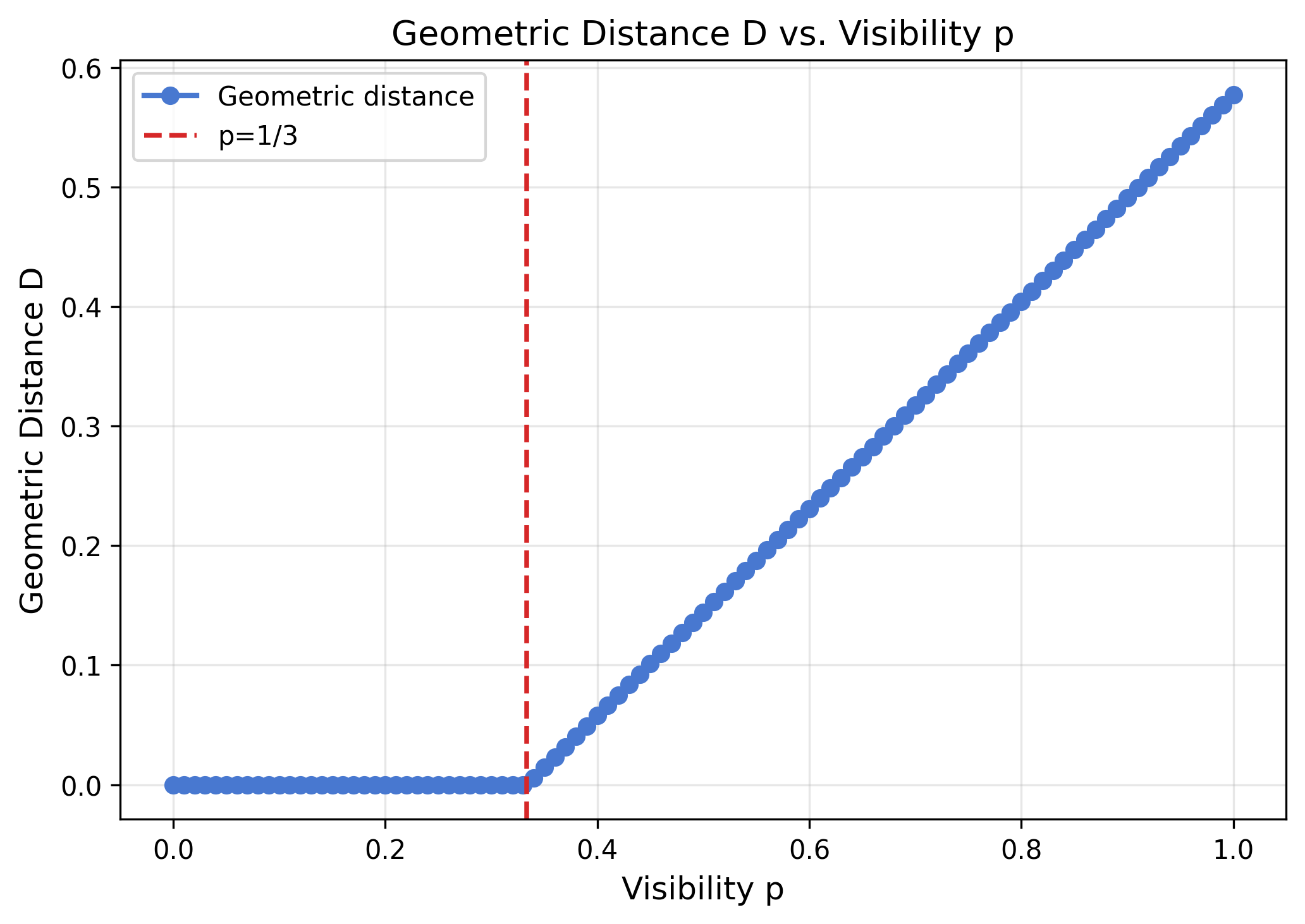}
    \caption{Geometric distance $D(\rho_{\rm W}(p))$ of the two-qubit
    Werner states. The numerical transition coincides with the exact
    teleportation threshold $p=1/3$.}
    \label{fig:example2}
\end{figure}

Therefore, $F_{\max}[\rho_{\rm W}(p)]>1/2$ if and only if $p>1/3$,
and the exact teleportation threshold is $p_{\rm tel}=1/3$. As shown
in Fig.~\ref{fig:example2}, the geometric distance is numerically zero
for $p\leq1/3$ and becomes positive for $p>1/3$, in agreement with the
analytical threshold.

For all sampled values of $p>1/3$, the closest state is numerically
found to be the boundary state $\rho_{\rm W}(1/3)$. Hence the
displacement from the target state to the boundary has a fixed
direction, and the projection-defined numerical witnesses are parallel
throughout the sampled teleportation-useful region. Under the
unit-Frobenius convention of Eq.~\eqref{eq:example_normalized_witness},
the common direction is
\begin{equation}
\widehat W_{\rm W}
=
\frac14
\left(
I_2\otimes I_2
-
\sigma_x\otimes\sigma_x
+
\sigma_y\otimes\sigma_y
-
\sigma_z\otimes\sigma_z
\right).
\label{eq:werner_witness}
\end{equation}
It satisfies $\|\widehat W_{\rm W}\|_F=1$ and
\begin{equation}
\operatorname{Tr}
[\widehat W_{\rm W}\rho_{\rm W}(p)]
=
\frac{1-3p}{4}.
\label{eq:werner_witness_expectation}
\end{equation}

Therefore, the witness becomes negative exactly for $p>1/3$.

The operator in Eq.~\eqref{eq:werner_witness} coincides with the
teleportation-witness form proposed in
Ref.~\cite{PhysRevLett.107.270501}. The closest-point construction
further identifies this direction as the geometrically optimal one for
the Werner family. This result is also consistent with related
teleportation-witness results in
Refs.~\cite{PhysRevA.86.032315,PhysRevA.87.042104}. For this particular
family, the entanglement threshold also occurs at $p=1/3$. Thus every
entangled state in this family is useful for teleportation, consistently
with Ref.~\cite{PhysRevA.87.042104}.

\subsubsection{Maximally Entangled Mixed States}

We next consider a rank-3 subclass of the two-qubit maximally entangled
mixed states introduced by Ishizaka \textit{et al.}
\cite{PhysRevA.62.022310},
\begin{equation}
\begin{aligned}
\rho_{\rm M}(p)
={}&
\frac{1+2p}{3}
|\psi^-\rangle\langle\psi^-|\\
&+
\frac{1-p}{3}
\left(
|00\rangle\langle00|
+
|\psi^+\rangle\langle\psi^+|
\right),
\qquad
p\in[0,1],
\end{aligned}
\label{eq:mems_state}
\end{equation}
where $|\psi^\pm\rangle=(|01\rangle\pm|10\rangle)/\sqrt2$.

Existing Bell-, steering-, and correlation-based criteria identify only
subsets of the teleportation-useful region. For this family, the
sufficient conditions are $p>(6\sqrt3-2)/13$ for the Bell-CHSH
criterion, $p>(3\sqrt5-1)/11$ for the three-setting linear steering
criterion, and $p>(3\sqrt5-1)/13$ for the average-correlation criterion
\cite{lmdx-dk95}.

\begin{figure}[htbp]
    \centering
    \includegraphics[width=\columnwidth]{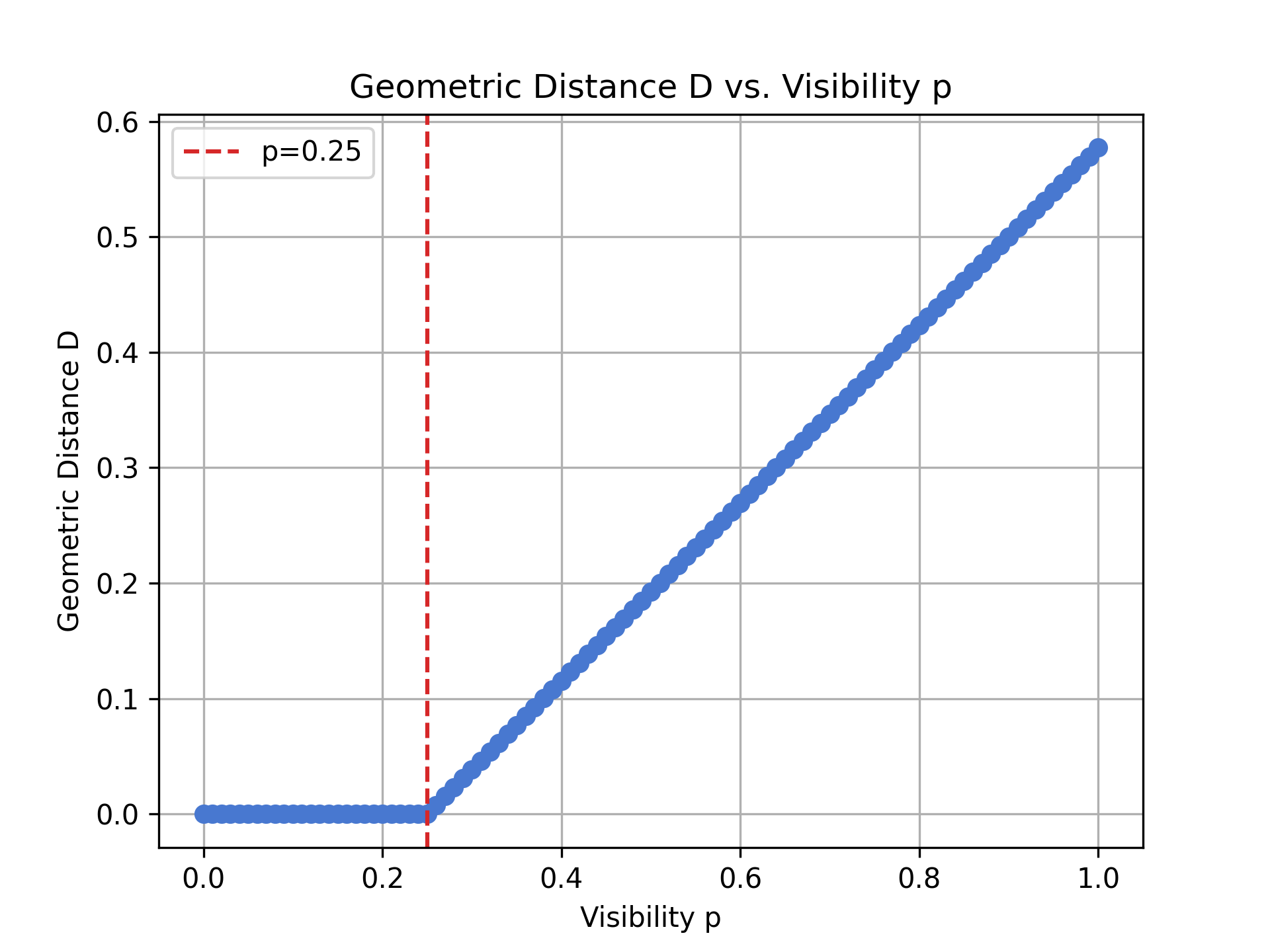}
    \caption{Geometric distance $D(\rho_{\rm M}(p))$ of the rank-3
    maximally entangled mixed states. The numerical transition coincides
    with the exact threshold $p=1/4$.}
    \label{fig:example1}
\end{figure}

In contrast, the geometric distance becomes positive at $p=1/4$, as
shown in Fig.~\ref{fig:example1}. This transition can be verified
analytically from the optimal teleportation fidelity. Define the
two-qubit correlation matrix by
\begin{equation}
T_{ij}
=
\operatorname{Tr}
[\rho(\sigma_i\otimes\sigma_j)].
\label{eq:correlation_matrix}
\end{equation}
For $\rho_{\rm M}(p)$,
\begin{equation}
T
=
\operatorname{diag}
\left(
-p,-p,-\frac{1+2p}{3}
\right).
\label{eq:mems_correlation_matrix}
\end{equation}
Since $\det T\leq0$ for $p\in[0,1]$, the optimal teleportation fidelity
for this family can be written as \cite{HORODECKI199621}
\begin{equation}
F(\rho)
=
\frac12
\left[
1+\frac13N(\rho)
\right],
\label{eq:twoqubit_fidelity}
\end{equation}
where $N(\rho)=\operatorname{Tr}\sqrt{T^\dagger T}$. For the present
family,
\begin{equation}
N[\rho_{\rm M}(p)]
=
2p+\frac{1+2p}{3}
=
\frac{1+8p}{3}.
\label{eq:mems_N}
\end{equation}

Substituting Eq.~\eqref{eq:mems_N} into Eq.~\eqref{eq:twoqubit_fidelity}
gives
\begin{equation}
F[\rho_{\rm M}(p)]
=
\frac{5+4p}{9}.
\label{eq:mems_fidelity}
\end{equation}
Hence,
\begin{equation}
F[\rho_{\rm M}(p)]>\frac23
\iff
p>\frac14,
\label{eq:mems_threshold}
\end{equation}
which agrees exactly with the transition obtained from the geometric
distance.

The numerical projections over the sampled useful region show that,
although the closest boundary point varies with $p$, the displacement
from $\rho_{\rm M}(p)$ to the boundary retains the same direction.
Consequently, the extracted projection-defined numerical witnesses are
parallel over the sampled values. Their common unit-Frobenius direction
is
\begin{equation}
\widehat W_{\rm M}
=
\frac14
\left(
I_2\otimes I_2
+
\sigma_x\otimes\sigma_x
+
\sigma_y\otimes\sigma_y
+
\sigma_z\otimes\sigma_z
\right).
\label{eq:mems_witness}
\end{equation}

This operator satisfies $\|\widehat W_{\rm M}\|_F=1$, and its
expectation value is
\begin{equation}
\operatorname{Tr}
[\widehat W_{\rm M}\rho_{\rm M}(p)]
=
\frac{1-4p}{6}.
\label{eq:mems_witness_expectation}
\end{equation}

Thus, $\operatorname{Tr}[\widehat W_{\rm M}\rho_{\rm M}(p)]<0$ if and
only if $p>1/4$. Therefore, within this family, $\widehat W_{\rm M}$
detects exactly the teleportation-usefulness transition. The
closest-point construction further identifies this fixed direction as
the geometrically optimal supporting-witness direction associated with
the exact boundary.

\subsection{Two-Qutrit Systems}

The preceding two-qubit examples exhibit fixed optimal witness
directions throughout their teleportation-useful regions. We next
consider an asymmetric two-qutrit family for which the closest boundary
point and the corresponding optimal witness direction vary with the
state parameter.

Consider the nonmaximally entangled state
\begin{equation}
|\Psi_{\mathrm{NME}}\rangle
=
\sqrt{\frac23}\,|00\rangle
+
\sqrt{\frac16}\,|11\rangle
+
\sqrt{\frac16}\,|22\rangle,
\label{eq:qutrit_nme}
\end{equation}
and the separable state
\begin{equation}
\sigma_+
=
\frac13
\left(
|01\rangle\langle01|
+
|12\rangle\langle12|
+
|20\rangle\langle20|
\right)
\label{eq:qutrit_sigma_plus}
\end{equation}
\cite{PhysRevLett.82.1056}. We define
\begin{equation}
\rho_{\rm Q}(p)
=
p|\Psi_{\mathrm{NME}}\rangle
\langle\Psi_{\mathrm{NME}}|
+
(1-p)\sigma_+,
\qquad
p\in[0,1].
\label{eq:qutrit_state}
\end{equation}

Let $p_c$ denote the critical teleportation threshold of this family.

\begin{figure}[htbp]
    \centering
    \includegraphics[width=\columnwidth]{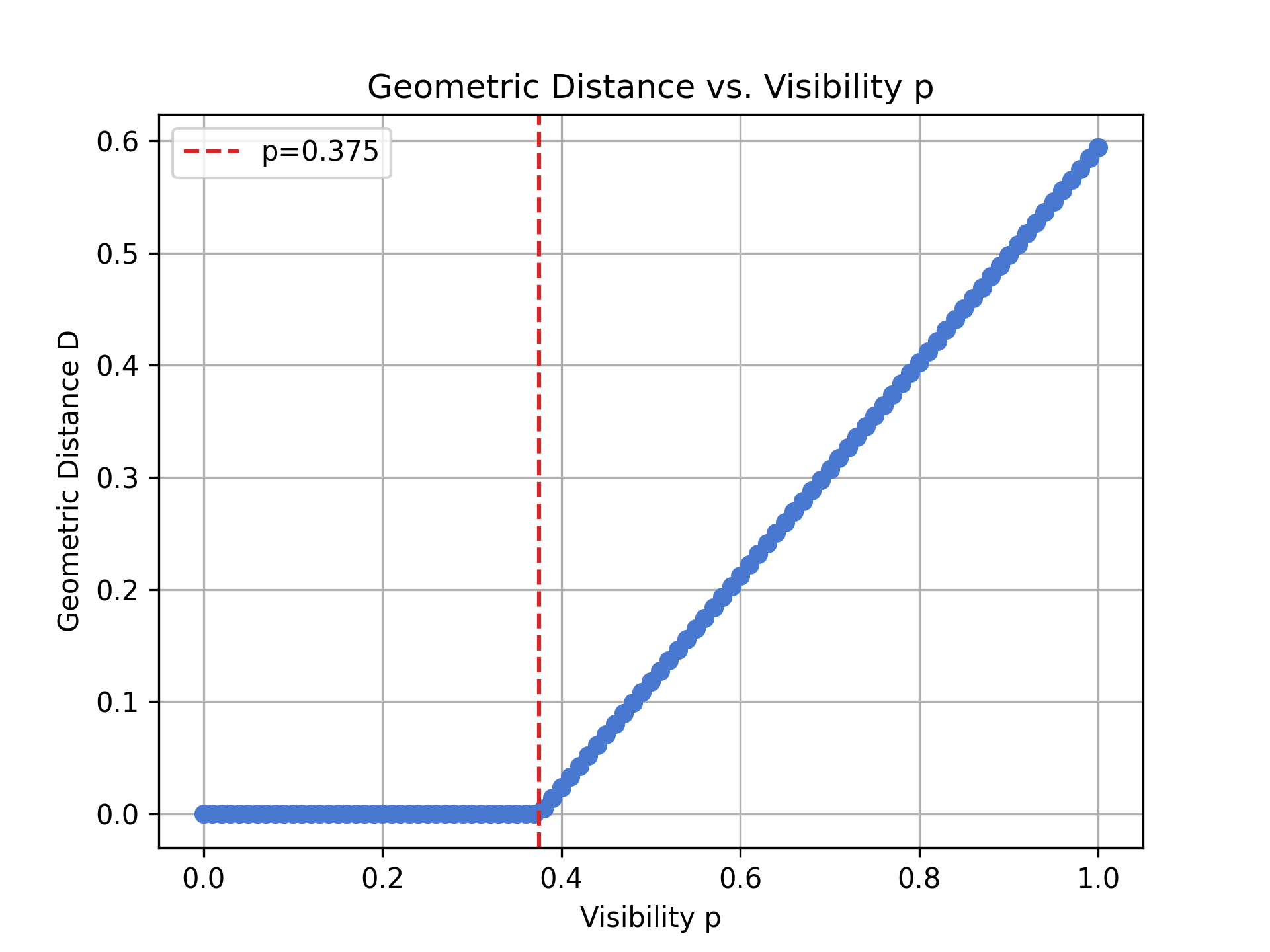}
    \caption{Geometric distance $D(\rho_{\rm Q}(p))$ of the asymmetric
    two-qutrit family. The numerical onset occurs near $p=3/8$,
    coinciding with the analytical sufficient bound in
    Eq.~\eqref{eq:qutrit_mef_bound}.}
    \label{fig:example3}
\end{figure}

As shown in Fig.~\ref{fig:example3}, the geometric distance is
numerically indistinguishable from zero at small $p$ and becomes
positive near $p=3/8$. An analytical sufficient condition follows from
the maximal entangled fraction. Choosing $U=I_3$ gives
\begin{equation}
F_{\max}[\rho_{\rm Q}(p)]
\geq
\langle\Phi_3^+|
\rho_{\rm Q}(p)
|\Phi_3^+\rangle
=
\frac{8p}{9},
\label{eq:qutrit_mef_bound}
\end{equation}
where
\begin{equation}
|\Phi_3^+\rangle
=
\frac1{\sqrt3}
\left(
|00\rangle+|11\rangle+|22\rangle
\right).
\label{eq:qutrit_phi}
\end{equation}

Hence $p>3/8$ is sufficient for teleportation usefulness, implying
$p_c\leq3/8$. The numerical onset of a positive distance occurs at the
same value within the adopted resolution, providing strong numerical
evidence that $p_c=3/8$ for this family. Equation
\eqref{eq:qutrit_mef_bound} alone, however, establishes only the
sufficient direction.

For convenience, denote the exact closest point corresponding to
$\rho_{\rm Q}(p)$ by $\sigma_p^*\equiv\sigma^*[\rho_{\rm Q}(p)]$. The
numerical projections indicate that this closest boundary point moves
with $p$, and the corresponding displacement does not retain a fixed
direction. Hence the extracted normalized numerical witness
$\widehat W_{\mathrm{out}}[\rho_{\rm Q}(p)]$ is parameter dependent.

To display the corresponding Gell-Mann coefficients on a convenient
numerical scale, we introduce only for this purpose the positively
rescaled representative
\begin{equation}
\widetilde W_{\rm Q}(p)
=
\frac{
9W_{\mathrm{out}}[\rho_{\rm Q}(p)]
}{
\displaystyle
\max_{i,j}
\left|
\left(
W_{\mathrm{out}}[\rho_{\rm Q}(p)]
\right)_{ij}
\right|
}.
\label{eq:qutrit_witness_scaling}
\end{equation}

This display rescaling does not affect the witness direction or its
detection condition. We expand $\widetilde W_{\rm Q}(p)$ in the
Gell-Mann tensor-product basis
$\{I_3,\lambda_1,\ldots,\lambda_8\}^{\otimes2}$, with
$\operatorname{Tr}(\lambda_i\lambda_j)=2\delta_{ij}$
\cite{PhysRev.125.1067}.

To illustrate the variation of the witness direction, we consider the
pure-state limit $p=1$ and a point just above the transition,
$p=0.376$. For $p=1$, the displayed representative is
{\small
\begin{equation}
\begin{aligned}
\widetilde W_{\rm Q}(1)\simeq{}&
5.9695\,I_3\otimes I_3\\
&-0.2123
\left(
I_3\otimes\lambda_3+\lambda_3\otimes I_3
\right)\\
&-0.1226
\left(
I_3\otimes\lambda_8+\lambda_8\otimes I_3
\right)\\
&-4.5000
\left(
\lambda_1\otimes\lambda_1+
\lambda_4\otimes\lambda_4
\right)\\
&+4.5000
\left(
\lambda_2\otimes\lambda_2+
\lambda_5\otimes\lambda_5
\right)\\
&-3.7761\,\lambda_3\otimes\lambda_3
-3.5638\,\lambda_8\otimes\lambda_8\\
&-0.1838
\left(
\lambda_3\otimes\lambda_8+
\lambda_8\otimes\lambda_3
\right)\\
&-3.4577\,\lambda_6\otimes\lambda_6
+3.4577\,\lambda_7\otimes\lambda_7.
\end{aligned}
\label{eq:qutrit_witness_p1}
\end{equation}
}

The single-party terms and nonuniform two-party coefficients reflect the
asymmetric structure of $|\Psi_{\mathrm{NME}}\rangle$.

At $p=0.376$, the displayed representative becomes
{\small
\begin{equation}
\begin{aligned}
\widetilde W_{\rm Q}(0.376)\simeq{}&
6.0000\,I_3\otimes I_3\\
&-4.5000\,\lambda_1\otimes\lambda_1
+4.5000\,\lambda_2\otimes\lambda_2\\
&-4.4979\,\lambda_3\otimes\lambda_3
-4.5000\,\lambda_4\otimes\lambda_4\\
&+4.5000\,\lambda_5\otimes\lambda_5
-4.4970\,\lambda_6\otimes\lambda_6\\
&+4.4970\,\lambda_7\otimes\lambda_7
-4.4975\,\lambda_8\otimes\lambda_8.
\end{aligned}
\label{eq:qutrit_witness_p0376}
\end{equation}
}

Near the transition, the single-party terms are numerically negligible
at the adopted reporting threshold, while the dominant two-party
coefficients have nearly uniform magnitudes. The contrast between
Eqs.~\eqref{eq:qutrit_witness_p1} and
\eqref{eq:qutrit_witness_p0376} shows that the extracted supporting
direction changes as the closest boundary point moves with $p$.

Unlike the two-qubit Werner and maximally entangled mixed-state
families, this asymmetric two-qutrit family therefore exhibits a
parameter-dependent projection-defined numerical witness direction.
This example illustrates the ability of the closest-point construction
to resolve nontrivial changes in the supporting geometry.

\subsection{Higher-Dimensional Systems}

We finally examine the method in higher-dimensional systems using
special generalized Werner states (SGWSs) with local dimensions
$d=4,\ldots,8$. This family admits an analytical maximal entangled
fraction and hence an exact teleportation threshold, allowing a direct
comparison with the numerically obtained geometric distance. It also
provides a convenient setting for examining how the optimal witness
direction changes away from the teleportation boundary.

The SGWS family introduced by Deng and Chen
\cite{Deng2009SGWS} is, in the bipartite case considered here,
\begin{equation}
\rho_d(p)
=
(1-p)\frac{I_{d^2}}{d^2}
+
p|\psi_d\rangle\langle\psi_d|,
\qquad
p\in[0,1],
\label{eq:sgws}
\end{equation}
where
\begin{equation}
|\psi_d\rangle
=
\sum_{i=0}^{d-1}\alpha_i|ii\rangle.
\label{eq:sgws_pure}
\end{equation}

The coefficients satisfy $\sum_{i=0}^{d-1}|\alpha_i|^2=1$.
Deng and Chen further considered the one-parameter subclass
\begin{equation}
\alpha_0
=
\frac{\cos\theta}{\sqrt d},
\qquad
\alpha_i
=
\sqrt{
\frac{d-\cos^2\theta}{d(d-1)}
},
\quad
i=1,\ldots,d-1.
\label{eq:sgws_theta}
\end{equation}
We choose $\theta=\pi/4$, giving
\begin{equation}
\alpha_0
=
\frac1{\sqrt{2d}},
\qquad
\alpha_i
=
\sqrt{
\frac{2d-1}{2d(d-1)}
},
\quad
i=1,\ldots,d-1.
\label{eq:sgws_coefficients}
\end{equation}

For the pure state in Eq.~\eqref{eq:sgws_pure},
\begin{equation}
F_{\mathrm{pure}}^{(d)}
=
\frac1d
\left(
\sum_{i=0}^{d-1}\alpha_i
\right)^2
=
\frac{
\left[
1+\sqrt{(d-1)(2d-1)}
\right]^2
}{
2d^2
}.
\label{eq:sgws_pure_mef}
\end{equation}

Since the maximally mixed component contributes $1/d^2$ to every
maximally entangled projector,
\begin{equation}
F_{\max}[\rho_d(p)]
=
\frac{1-p}{d^2}
+
pF_{\mathrm{pure}}^{(d)}.
\label{eq:sgws_mef}
\end{equation}
The exact teleportation threshold obtained from $F_{\max}=1/d$ is
therefore
\begin{equation}
p_{\mathrm{tel}}^{(d)}
=
\frac{
\frac1d-\frac1{d^2}
}{
F_{\mathrm{pure}}^{(d)}-\frac1{d^2}
}.
\label{eq:sgws_threshold}
\end{equation}

The resulting thresholds are listed in
Table~\ref{tab:sgws_thresholds}.

\begin{table}[htbp]
\centering
\caption{Analytical teleportation thresholds
$p_{\mathrm{tel}}^{(d)}$ for the SGWS family.}
\label{tab:sgws_thresholds}
\begin{tabular}{c|ccccc}
\hline
$d$
& $4$
& $5$
& $6$
& $7$
& $8$\\
\hline
$p_{\mathrm{tel}}^{(d)}$
& $0.205725$
& $0.170213$
& $0.145280$
& $0.126765$
& $0.112455$\\
\hline
\end{tabular}
\end{table}

\begin{figure}[htbp]
    \centering
    \includegraphics[width=\columnwidth]{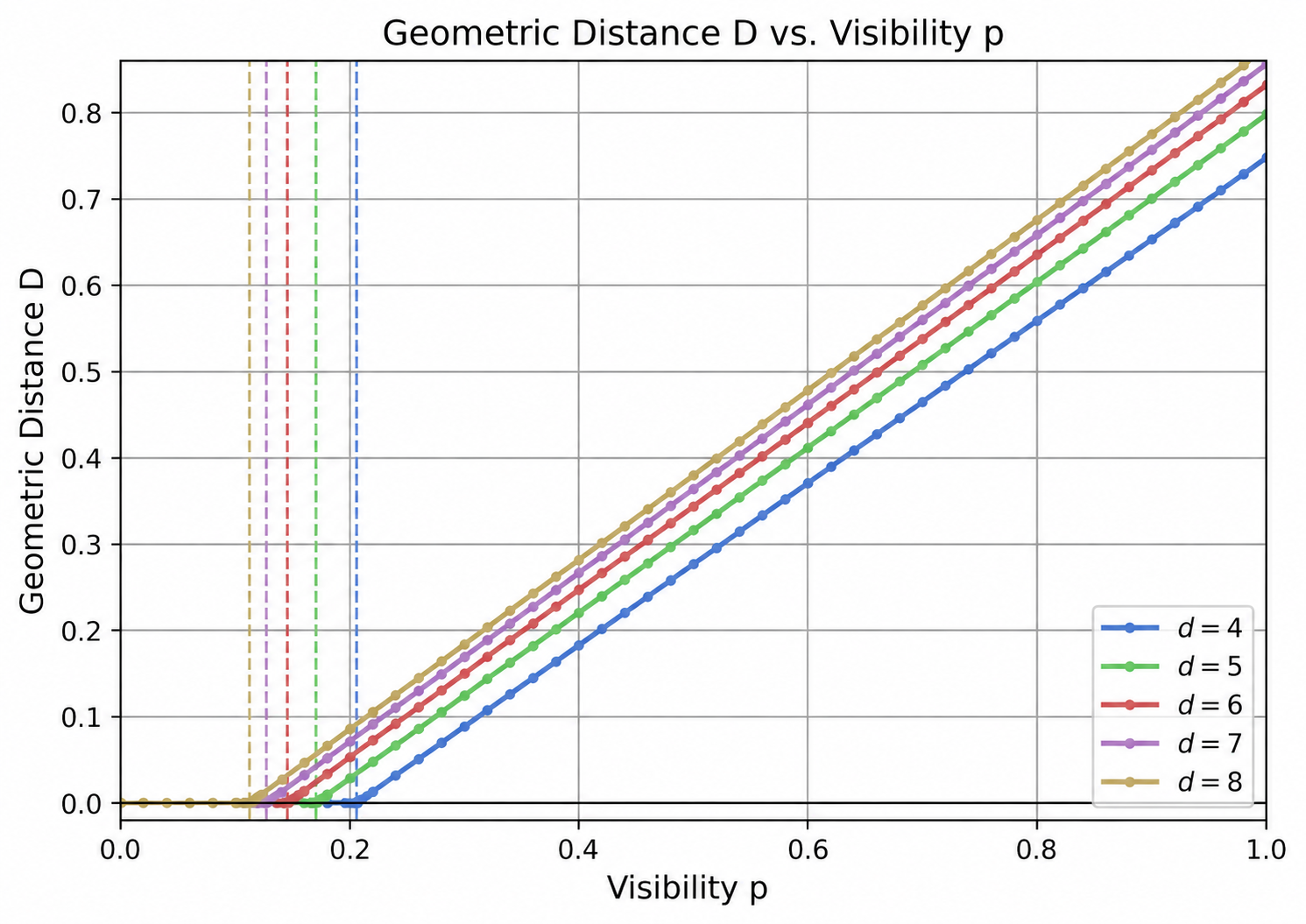}
    \caption{Geometric distance $D(\rho_d(p))$ of the SGWS family for
    $d=4,5,6,7,$ and $8$. The dashed vertical lines indicate the
    corresponding analytical thresholds $p_{\mathrm{tel}}^{(d)}$.}
    \label{fig:example4}
\end{figure}

We apply the same two-layer iterative cutting-plane method for
$d=4,\ldots,8$. As shown in Fig.~\ref{fig:example4}, the geometric
distance is numerically indistinguishable from zero below the
corresponding analytical threshold and becomes positive when the
threshold is crossed. The numerical transitions agree with
Table~\ref{tab:sgws_thresholds} for all dimensions considered.

For all dimensions considered, the projection-defined optimal witness
obtained just above the corresponding teleportation threshold is
numerically parallel to the direction
\begin{equation}
W_{d,\mathrm{th}}
=
\frac1d I_{d^2}
-
|\Phi_d^+\rangle\langle\Phi_d^+|.
\label{eq:sgws_threshold_witness}
\end{equation}

Here $|\Phi_d^+\rangle=\frac1{\sqrt d}\sum_{j=0}^{d-1}|jj\rangle$ is
the $d$-dimensional maximally entangled state.
The same operator form has previously appeared as a teleportation
witness in Ref.~\cite{PhysRevLett.107.270501}. Here it is used only to
identify the threshold-witness direction selected numerically by the
closest-point construction.

Its Frobenius norm is
$\|W_{d,\mathrm{th}}\|_F=\sqrt{2(d-1)/d}$, and the corresponding
unit-Frobenius representative is
\begin{equation}
\widehat W_{d,\mathrm{th}}
=
\sqrt{\frac{d}{2(d-1)}}
\left(
\frac1d I_{d^2}
-
|\Phi_d^+\rangle\langle\Phi_d^+|
\right).
\label{eq:sgws_normalized_threshold_witness}
\end{equation}

We denote the unit-Frobenius numerical witness obtained for
$\rho_d(p)$ by
$\widehat W_d(p)\equiv\widehat W_{\mathrm{out}}[\rho_d(p)]$.

To quantify the change in witness direction, we define the normalized
Hilbert--Schmidt overlap
\begin{equation}
\Gamma_d(p)
=
\operatorname{Tr}
\left[
\widehat W_d(p)
\widehat W_{d,\mathrm{th}}
\right].
\label{eq:sgws_witness_overlap}
\end{equation}

Because both operators have unit Frobenius norm,
$-1\leq\Gamma_d(p)\leq1$. In particular, $\Gamma_d(p)=1$ means that
$\widehat W_d(p)=\widehat W_{d,\mathrm{th}}$, i.e., the two witnesses
have the same orientation and define the same supporting-hyperplane
direction.

For convenience, we define
\begin{equation}
\Delta_d(p)
=
1-\Gamma_d(p)
\label{eq:sgws_witness_deviation}
\end{equation}
as the directional deviation from the threshold-witness direction.
Thus, $\Delta_d(p)=0$ corresponds to identical directions, whereas a
larger value indicates a larger departure. Representative values are
listed in Table~\ref{tab:sgws_witness_deviation}.

\begin{table}[htbp]
\centering
\caption{Directional deviation
$\Delta_d(p)=1-\Gamma_d(p)$ from the normalized threshold-witness
direction $\widehat W_{d,\mathrm{th}}$ for representative values of
$p$. Values at the $10^{-12}$ level are numerically indistinguishable
from zero.}
\label{tab:sgws_witness_deviation}
\begin{tabular}{c|cccc}
\hline
$d$
& $p=0.6$
& $p=0.8$
& $p=0.9$
& $p=1$\\
\hline
$4$
& $4.52\times10^{-13}$
& $3.48\times10^{-13}$
& $5.53\times10^{-5}$
& $3.84\times10^{-4}$\\
$5$
& $1.08\times10^{-12}$
& $3.02\times10^{-4}$
& $1.04\times10^{-3}$
& $1.88\times10^{-3}$\\
$6$
& $5.76\times10^{-6}$
& $1.22\times10^{-3}$
& $1.98\times10^{-3}$
& $2.67\times10^{-3}$\\
$7$
& $4.04\times10^{-4}$
& $1.81\times10^{-3}$
& $2.43\times10^{-3}$
& $2.96\times10^{-3}$\\
$8$
& $8.59\times10^{-4}$
& $2.11\times10^{-3}$
& $2.60\times10^{-3}$
& $3.02\times10^{-3}$\\
\hline
\end{tabular}
\end{table}

Near the teleportation boundary, the extracted numerical witness is
parallel to $W_{d,\mathrm{th}}$. As $p$ increases, the
projection-defined numerical direction gradually departs from this
threshold form. The deviation becomes more pronounced toward the
pure-state limit and is generally larger for the higher dimensions
considered here. Thus, the closest-point construction provides
information not only about the teleportation threshold, but also about
the parameter dependence of the optimal supporting direction.

Finally, we examine the computational cost of the higher-dimensional
calculations. All calculations were performed on a computational node
equipped with 16 Intel Xeon Gold 6430 vCPUs and 120~GB of RAM. For
representative calculations at $p=1$, the runtime for a single state
increased from approximately $0.86~\mathrm{s}$ at $d=4$ to
approximately $204~\mathrm{s}$ at $d=8$. This increase is consistent
with the outer quadratic semidefinite optimization becoming the
dominant computational bottleneck as the local dimension grows.
Nevertheless, all cases up to $d=8$ remained directly tractable within
the same numerical framework.

\section{Conclusion}

In this work, we present a distance-based geometric method for
constructing optimal teleportation witnesses for arbitrary two-qudit
states, which can be used to determine whether a given state $\rho$ is
useful for teleportation. By solving the shortest-distance problem from
$\rho$ to the closed convex set $S$ of states that are not useful for
teleportation, we obtain the unique projection point $\sigma^*\in S$
and construct the corresponding optimal teleportation witness from the
supporting hyperplane at $\sigma^*$. The resulting distance is positive
if and only if $\rho$ is useful for teleportation. Since this
shortest-distance problem involves infinitely many constraints
parameterized by the continuous unitary group $\mathrm{U}(d)$, we
develop a two-layer cutting-plane method to solve the resulting convex
semi-infinite problem numerically and approximate $\sigma^*$. We further validate the proposed method on four representative classes of bipartite states, demonstrating its applicability from two-qubit systems to higher-dimensional bipartite states.

Closest-point constructions of optimal witnesses have also been studied in entanglement theory, where entanglement witnesses can be derived from the closest separable state
\cite{VedralPlenio1998,WITTE199914,PhysRevA.98.012315,
PhysRevLett.120.050506,PhysRevA.92.042322,
PhysRevA.89.052319,PhysRevA.80.032324,Hayashi2009geometric,
PhysRevA.67.012327,PhysRevA.66.032319}. Although our construction is based on the same closest-point geometry, the numerical methods developed for the separable-state problem cannot be directly transferred to the present setting, because the projection is taken onto the teleportation-useless set $S$. Existing variational, neural-network, and Gilbert-type approaches exploit the convex-mixture structure of separable states in terms of product states
\cite{PhysRevA.106.062413,PhysRevResearch.4.023238,PhysRevA.102.012409}. In contrast, the teleportation-useless set is naturally characterized as the intersection of infinitely many unitary-parameterized half-spaces, so its closest-point problem takes the form of a convex semi-infinite optimization problem
\cite{doi:10.1137/1035089}. This distinct structure underlies the cutting-plane treatment adopted here. In practice, the nonconvex inner unitary optimization is solved numerically by a randomized multi-start fixed-point iteration, so the accuracy of the final approximation depends on the numerical search budget and prescribed tolerances.

Overall, our geometric construction method provides a general and operational framework for studying the teleportation usefulness of arbitrary two-qudit entangled states. For multipartite entangled states \cite{mv3y-btlq, Yonezawa2004,PhysRevLett.122.170501}, once the geometric structure of the set of states useless for teleportation can be characterized through an analogous relation between teleportation fidelity and the MEF, our framework may be naturally extended to study the teleportation usefulness of multipartite entangled states.

\section*{ACKNOWLEDGMENT}
This work is supported by the National Natural Science Foundation of China (Grants No. 62571060, No. 62171056, and No. 62220106012).
\bibliographystyle{apsrev4-2}  % 选择引用格式，revtex4-2 推荐 apsrev4-2
\bibliography{ref}            % ref.bib 文件名（不加 .bib）

\end{document}